\documentclass[english,aps,preprint]{revtex4}
\usepackage[T1]{fontenc}
\usepackage[latin1]{inputenc}
\usepackage{babel}
\usepackage{graphics}

\makeatletter

\providecommand{\LyX}{L\kern-.1667em\lower.25em\hbox{Y}\kern-.125emX\@}

\makeatother
\begin{document}
\hspace*{10cm} COLO-HEP-464\\
 \hspace*{10.5cm} May 2001.

\noindent \vspace*{1cm}

{\centering \textbf{\Large Topological Susceptibility on Dynamical
Staggered Fermion Configurations}\Large \par}

{\centering \vspace{0.5cm}\par}

{\centering {\large Anna Hasenfratz\( ^{\dagger } \) }\large \par}

{\centering \vspace{0.5cm}\par}

{\centering Physics Department, University of Colorado, \\
 Boulder, CO 80309 USA\par}

{\centering \vspace{0.5cm}\par}

{\centering \textbf{Abstract}\par}

{\centering \vspace{0.5cm}\par}

{\centering The topological susceptibility is one of the few physical
quantities that directly measure the properties of the QCD vacuum.
Chiral perturbation theory predicts that in the small quark mass limit
the topological susceptibility depends quadratically on the pion mass,
approaching zero in the chiral limit. Lattice calculations have difficulty
reproducing this behavior. In this paper we study the topological
susceptibility on dynamical staggered fermion configurations. Our
results indicate that the lattice spacing has to be small, around
\( a\approx 0.1 \)fm for thin link staggered fermion actions to show
the expected chiral behavior. Our preliminary result indicates that
fat link fermions, on the other hand, reproduce the theoretical expectations
even on lattices with \( a\approx 0.17 \)fm. We argue that this is
due to the improved flavor symmetry of fat link fermionic actions. \par}

\vspace{0.5cm}


\vfill

\( ^{\dagger } \) {\small e-mail: anna@eotvos.colorado.edu}{\small \par}

\section{Introduction}

Instantons play an important role in the QCD vacuum from the breaking
of the axial \( U_{A}(1) \) symmetry to chiral symmetry breaking
and the low energy hadron spectrum\cite{Schafer:1998wv-shuryak-revmodphys}.
Lattice studies support many of the theoretical predictions\cite{GarciaPerez:2000hq-lattice-proc}.
For example, the Witten-Veneziano formula relates the pure gauge topological
susceptibility to the masses of the \( \eta ,\eta ' \) and K mesons
predicting \( \chi _{\infty }^{1/4}=180 \)MeV. The topological susceptibility
on pure gauge lattices has been measured by several groups using different
methods obtaining consistent results \( \chi ^{1/4}_{\infty }=205(5) \)MeV\cite{Hasenfratz:1998qk-SU3_topo., deForcrand:1998yw-full_QCD_topo, Alles:1997nm-pure-gauge-SU(3)}. 

Light fermions suppress the topological susceptibility. Chiral perturbation
theory predicts\cite{Leutwyler:1992yt} \begin{equation}
\label{chi dynamical}
\chi =\frac{m_{q}}{n_{f}}\Sigma +O(m_{q}^{2})=\frac{f_{\pi }^{2}m^{2}_{\pi }}{2n_{f}}+O(m^{4}_{\pi }),
\end{equation}
 where \( m_{q} \) is the quark mass, \( n_{f} \) is the number
of fermion flavors and \( \Sigma  \) is the chiral condensate per
fermionic flavor. In this normalization \( f_{\pi }=92 \)MeV. Several
recent lattice studies measured \( \chi  \) on dynamical configurations.
Calculations with Wilson and clover fermions cover a fair range of
lattice spacing \( a\approx 0.08-0.20 \)fm and pion masses \( m_{\pi }r_{0}\approx 1.3-2.5 \)
(\( r_{0} \) is the Sommer parameter \cite{Sommer:1994ce-Sommer_r0})
\cite{Hart:1999hy-Wilson_topo, Hart:2000gh-Wils-top, Hart:2000wr-Wilson_top, Bali:2001gk-Wils_topo, AliKhan:1999zi}.
The results appear controversial. UKQCD uses clover fermions\cite{Hart:1999hy-Wilson_topo, Hart:2000gh-Wils-top, Hart:2000wr-Wilson_top}.
Their data on lattices with \( a\approx 0.1 \)fm is basically consistent
with eqn.(\ref{chi dynamical}). SESAM/T\( \chi L \) uses unimproved
Wilson fermions. Their topological susceptibility at similar lattice
spacing does not decrease with the pion mass though the statistical
errors are too large to claim inconsistency with theoretical expectations\cite{Bali:2001gk-Wils_topo}.
CP-PACS published data using clover fermions and improved gauge action
at lattice spacing \( a\approx 0.17 \)fm\cite{AliKhan:1999zi}. Their
conclusion is the same as SESAM/T\( \chi L \). The situation with
staggered fermions is not much better. Only the Pisa group measured
the topological susceptibility with two and four flavors of staggered
fermions at lattice spacing \( a\approx 0.09-0.17 \)fm\cite{Alles:2000cg-nf=2-4, Alles:1999kf-nf4_staggered}.
They do not see the reduction of the topological susceptibility with
decreasing quark mass either.

Can we understand what is going on with the dynamical simulations?
Eqn.(\ref{chi dynamical}) is valid only in case of \( n_{f} \) light
\emph{chiral} fermions creating \( n_{f}^{2}-1 \) light pions. Both
Wilson and staggered fermions violate this assumption. Wilson fermions
break chiral symmetry explicitly. The addition of the clover term
reduces the symmetry breaking and improves chiral behavior. Staggered
fermions have only a residual \( U(1) \) chiral symmetry and only
one true Goldstone boson, the other pion-like states can be heavy.
Fat link fermions considerably improve flavor symmetry and consequently
chiral behavior. Both fermionic formulations should show the expected
chiral behavior in the continuum limit, clover and fat link fermions
sooner than the unimproved ones, but it is not clear when scaling
in the topological susceptibility actually sets in. In a recent paper
\cite{Durr:2001ty} it was suggested that the reason the topological
susceptibility from available lattice data does not follow the expected
theoretical behavior is a combination of three effects: too large
lattice spacing, too small volumes and too small Leutwyler-Smilga
\cite{Leutwyler:1992yt} parameter. While all three conditions could
indeed be important, we feel that the non-chiral behavior of the fermionic
actions is the main cause of the problem. Chiral symmetry violation
of the fermionic action is a scaling violation effect and is covered
by the first condition of \cite{Durr:2001ty}. However, it is not
a universal quantity, it can strongly depend on the fermionic action.

Why are the topological properties of the vacuum important? Phenomenological
instanton models predict that the low energy hadron spectrum is governed
by the near-zero eigenvalue modes of the Dirac operator, which, in
turn, are related to instantons. If the fermion-instanton interaction
is different from the continuum one either because of chiral symmetry
breaking or other lattice artifacts, the Dirac spectrum and consequently
the low energy hadron spectrum can also be different. Recently, using
chiral symmetric overlap fermions on the lattice, we showed that with
light quarks the first few modes of the Dirac operator saturate the
low lying hadron propagators on quenched \( a\approx 0.12 \) lattices,
just as the phenomenological models predict\cite{DeGrand:2000gq-overlap_eigenmodes, DeGrand:2001pj-overlap-comment}.
In a subsequent publication contradictory conclusion was reached using
Wilson fermions\cite{Horvath:2001ir-against-instanton-dom.}. For
us that implies that Wilson fermions, at least at large to moderate
lattice spacings, interact differently with instantons than chirally
symmetric fermions. 

In this paper we investigate the topological susceptibility on both
\( n_{f}=2 \) and \( n_{f}=4 \) staggered dynamical fermion configurations.
The \( n_{f}=2 \) configurations are \( 16^{3}32 \) thin link staggered
fermion lattices created by the Columbia and MILC collaborations and
downloaded from the NERSC archive\cite{NERSC}. The \( n_{f}=4 \)
lattices are smaller, \( 8^{3}24 \) configurations used in the study
of fat link fermions in Ref. \cite{Knechtli:2000ku-dynamical_fatlinks}.
Two of the \( n_{f}=4 \) sets were generated with thin link staggered
fermions and one with N=3 level APE blocked fat link fermions. The
latter action has about an order of magnitude better flavor symmetry
than thin link actions at similar parameter values. The thin link
staggered fermion results with \( a\leq 0.1 \)fm are more or less
consistent with eqn.(\ref{chi dynamical}). The thin link \( a\approx 0.17 \)fm
data shows clear deviation, the topological susceptibility is consistent
with the quenched value, independent of the pion mass. On the other
hand, the fat link data at the same coarse lattice spacing is in perfect
agreement with eqn.(\ref{chi dynamical}), suggesting that improved
flavor/chiral symmetry indeed has a strong effect on the topology. 

To determine the topological charge of the configurations we used
a topological charge density operator constructed from hypercubic
blocked (HYP) fat links\cite{Hasenfratz:2001hp-HYP-blocking}. Hypercubic
blocking was introduced in a recent paper as an alternative to repeated
APE smearing\cite{Albanese:1987ds-APE_blocking}. Hypercubic blocking
mixes links only from the hypercubes that attach to a given link,
thus the fat link is very compact yet the configuration is as smooth
as after three levels of APE blocking. To avoid the distortion effect
of extended operators, we consider only one to three levels of hypercubic
blocked operators (HYP1, HYP2 and HYP3). We have calculated the additive
and multiplicative renormalization factors of the topological susceptibility
for these operators. After two to three levels of HYP blocking the
renormalization factors turn out to be consistent with their tree
level values and, after correction, all three HYP topological susceptibility
measurements are consistent.

In Sect. 2. we describe the hypercubic topological operator and illustrate
the measurement of the renormalization factors on pure gauge Wilson
plaquette configurations. Sect. 3 contains our results for two- and
four-flavor staggered fermions. Sect. 4 is a short discussion on flavor
symmetry and the summary of our results.

\section{Hypercubic Topological Charge Operator}

Most large scale simulations use pure gauge observables to measure
the topological susceptibility with some discretized version of the
continuum \( F\tilde{F} \) as the lattice charge density operator
\( q_{L}(x) \). The relation between the continuum and lattice topological
susceptibilities contains both a multiplicative and an additive renormalization
factor. The lattice charge density operator is related to the continuum
one through a multiplicative renormalization factor\begin{equation}
q_{L}(x)=Za^{2}q(x).
\end{equation}
 In addition the correlator of two topological density operators have
an additive correction term as well due to the mixing of \( q_{L}(x) \)
with other lattice operators\begin{equation}
q_{L}(x)q_{L}(0)=Z^{2}a^{4}q(x)q(0)+m(x),
\end{equation}
 thus the topological susceptibility on the lattice is \begin{equation}
\chi _{L}=\int q(x)q(0)d^{4}x=Z^{2}a^{4}\chi +M,
\end{equation}
\begin{equation}
M=\int m(x)d^{4}x.
\end{equation}
The renormalization factors Z and M depend on the lattice parameters
\( \beta  \) and \( m_{q} \). The renormalization constants in principle
can be determined non-perturbatively using the heating method proposed
by the Pisa group\cite{Christou:1996zn-heating_improved}. 

Due to lattice artifacts, mainly dislocations, on typical lattices
\( Z\cong 0.25 \) and \( M/\chi _{L}\cong 1 \) for thin link topological
density operators which makes the direct determination of \( \chi  \)
almost impossible. Local smearing or cooling removes most of these
lattice artifacts moving Z towards 1 and M to 0. Repeated smoothing
gives \( Z\cong 1 \) and \( M/\chi \ll 1 \) leading to \( \chi _{L}\cong a^{4}\chi  \).
Since the non-perturbative calculation of Z and M introduces statistical
and systematical errors, a topological density operator where the
renormalization constants are small or can be neglected is desirable.
However, repeated smearing and cooling methods have their drawbacks
as well. While removing vacuum fluctuations and lattice artifacts,
both methods remove topological objects, mainly small instantons and
close-by pairs. In addition the size of the remaining objects change
as well, as can be demonstrated by monitoring individual instantons
during the smearing process\cite{Hasenfratz:1998qk-SU3_topo.}.

\begin{figure}
{\centering \resizebox*{7cm}{!}{\includegraphics{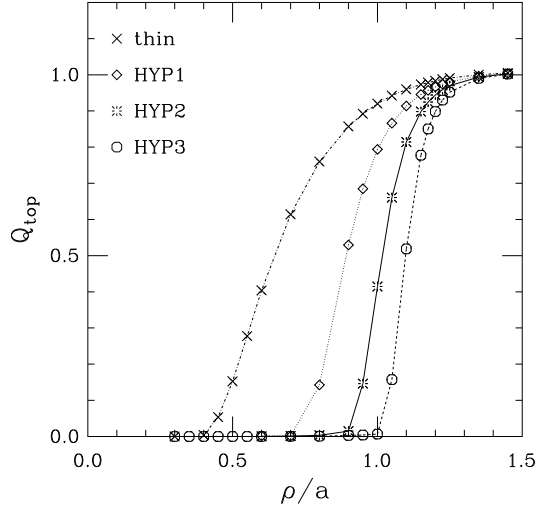}} \par}

\caption{\label{Q_vs_rho}The topological charge of a smooth instanton as
the function of its radius \protect\( \rho /a\protect \). The charge
is measured with thin link operator (crosses) , HYP1 (diamonds) ,
HYP2 (bursts) and HYP3 (octagons) fat link operators.}
\end{figure}
In this paper we construct the topological charge density operator
\( q_{L}(x) \) using hypercubic blocked links and the improved thin
link charge operator of Refs. \cite{DeGrand:1998ss-SU(2)_top-operator}
and \cite{Hasenfratz:1998qk-SU3_topo.}. Hypercubic blocking was introduced
in Ref. \cite{Hasenfratz:2001hp-HYP-blocking} as a localized alternative
to repeated APE blocking. HYP fat links mix original links from the
hypercubes that are attached to the fat link only, yet they create
configurations that are as smooth as the ones obtained after three
levels of APE blocking. 

Topological charge measurement are difficult because of the presence
of dislocations: short distance vacuum fluctuations that can be mistaken
for small instantons. Smearing the links in the topological density
operator reduces this problem in part by sharpening the transition
between the topologically non-trivial \( Q=\int d^{4}xq(x)=1 \) sector
and the trivial \( Q=0 \) sector. Figure \ref{Q_vs_rho} shows the
topological charge of a smooth instanton measured with the thin link,
HYP1, HYP2 and HYP3 fat link operators as the function of the instanton
radius . 
\begin{figure}
{\centering \resizebox*{14cm}{!}{\includegraphics{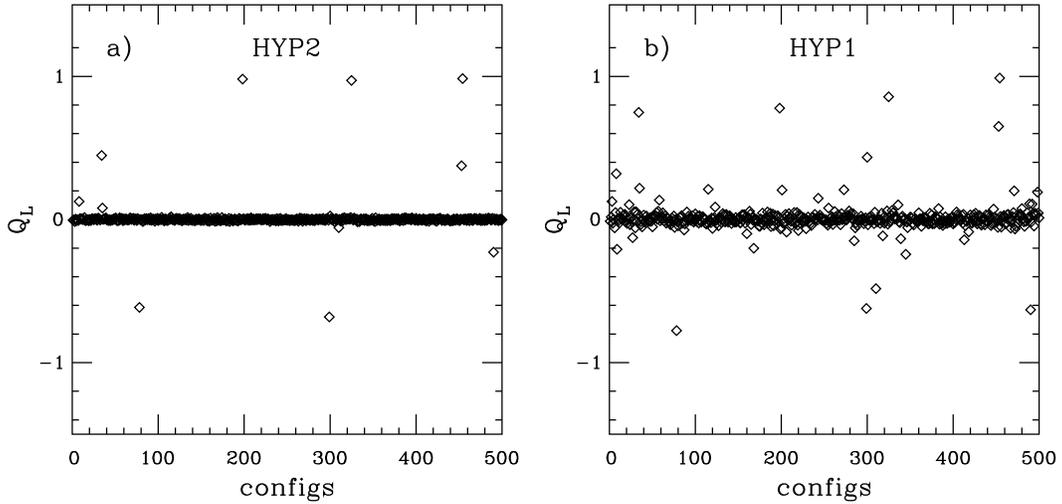}} \par}

\caption{\label{renorm.ps}The topological charge on 500 heated \protect\( 8^{4}\protect \)
\protect\( Q=0\protect \) configurations measured with a) HYP2 topological
operator, b) HYP1 topological operator. The configurations were heated
with 10-50 heat bath steps using \protect\( \beta =6.0\protect \)
Wilson pure gauge action.}
\end{figure}
The charge measured with the HYP operators rises sharply at instanton
radius \( \rho /a\approx 1 \). 

Even though the HYP operators remove most dislocations, they still
can have non-trivial renormalization factors. We have measured the
renormalization factors Z and M following the heating method of Ref.
\cite{Christou:1996zn-heating_improved}. To measure M one has to
heat a trivial \( Q=0 \) configuration. The configuration thermalizes
fast at short distances, and since the origin of M is local contact
terms, short distance thermalization is sufficient to measure M. Before
heating creates non-trivial topological objects the thermalization
must be terminated and restarted with a different random seed. It
is important to make sure the measurement of M is done on \( Q=0 \)
configurations only. Fortunately with the HYP3 and frequently with
the HYP2 operators it is easy to separate the trivial configurations
from the occasional \( Q\neq 0 \) ones. Figure \ref{renorm.ps}/a
shows the topological charge on 500 heated \( Q=0 \) configurations
measured with HYP2 operator. Ordered \( 8^{4} \) configurations were
heated with 10-50 heat bath steps using \( \beta =6.0 \) pure gauge
Wilson action. Most configurations have very small topological charge,
a few has \( |Q|\sim 1 \) and only 2-3 configurations have topological
charge whose interpretation is not clear. I chose, based on Figure
\ref{Q_vs_rho} but somewhat arbitrarily, a cut \( Q_{max}=0.3 \)
to separate the \( Q=0 \) and \( Q\neq 0 \) configurations. The
final results are fairly insensitive to the precise choice of this
cut. The topological charge measured with the HYP1 operator fluctuates
more as is illustrated in Figure \ref{renorm.ps}/b. Again, a configuration
is accepted as \( Q=0 \) if its topological charge is \( |Q_{L}|<0.3 \)
as measured with the HYP2 operator. The multiplicative renormalization
constant Z is calculated similarly on \( Q=1 \) configurations that
contain a single smooth instanton of size \( \rho /a=3.0,2.5 \) or
2.0. Z can be different on different instanton size backgrounds, though
in none of the measurements we performed was the difference more than
a few per cent. We chose Z on the background where the smooth instanton's
size was closest to the expected average instanton size of the Monte
Carlo configurations, \( \rho /a\approx 0.3 \)fm\( /a \). 
\begin{table}
{\centering \begin{tabular}{|c|c|c|c|c|c|}
\hline 
&
Z&
M\( \times 10^{5} \)&
\( \chi _{L}\times 10^{5} \)&
\( a^{4}\chi \times 10^{5} \)&
\( \chi r_{0}^{4} \)\\
\hline
\hline 
APE1&
0.795(4)&
0.95(7)&
5.4(5)&
7.0(7)&
0.058(6)\\
\hline 
HYP1&
0.935(4)&
0.11(3)&
6.3(4)&
7.1(5)&
0.059(6)\\
\hline 
HYP2&
1.000(4)&
0.006(4)&
7.0(4)&
7.0(4)&
0.058(6)\\
\hline
\end{tabular}\par}

\caption{\label{quenched_results_b6.0}Results for the renormalization factors
Z and M and the lattice and continuum topological susceptibilities
for \protect\( \beta =6.0\protect \) pure gauge Wilson action. The
topological susceptibility was measured on 220 \protect\( 16^{3}32\protect \)
configurations downloaded from the NERSC archive\cite{Kilcup:1997hp-NERSC_ref.}.}
\end{table}

\begin{table}
{\centering \begin{tabular}{|c|c|c|c|c|c|}
\hline 
&
Z&
M\( \times 10^{4} \)&
\( \chi _{L}\times 10^{4} \)&
\( a^{4}\chi \times 10^{4} \)&
\( \chi r_{0}^{4} \)\\
\hline
\hline 
HYP1&
0.85(4)&
0.6(1)&
5.1(9)&
6.2(14)&
0.048(12)\\
\hline 
HYP2&
0.95(1)&
0.19(4)&
6.0(9)&
6.4(12)&
0.049(11)\\
\hline 
HYP3&
0.98(1)&
0.02(1)&
6.2(9)&
6.4(10)&
0.049(9)\\
\hline
\end{tabular}\par}

\caption{\label{quenched_results_b5.7}Results for the renormalization factors
Z and M and the lattice and continuum topological susceptibilities
for the \protect\( \beta =5.7\protect \) pure gauge Wilson action.
The topological susceptibility was measured on 350 \protect\( 8^{3}24\protect \)
configurations.}
\end{table}

Results for \( \beta =6.0 \) pure gauge Wilson action are summarized
in Table \ref{quenched_results_b6.0} where Z, M, \( \chi _{L} \),
\( \chi a^{4}=(\chi _{L}-M)/Z^{2} \) and \( \chi r_{0}^{4} \) with
Sommer parameter \( r_{0}=5.37(1) \) are given both for the HYP1
and HYP2 operator, and, for comparison, one level APE smeared operator
with parameter \( \alpha =0.75 \) as well. The topological susceptibility
was measured on 220 \( 16^{3}32 \) \( \beta =6.0 \) configurations
from the NERSC archive\cite{Kilcup:1997hp-NERSC_ref.}. The results
for \( a^{4}\chi  \) are consistent for  all three operators, in
physical units \( \chi ^{1/4}=196(5) \)MeV. The renormalization factors,
on the other hand, are quite different. The background term M is about
17\% of \( \chi _{L} \) for the APE1 operator, 2\% for the HYP1 operator
and less than 0.1\% for the HYP2 operator. The renormalization factors
for the HYP2 operator are negligible, for all practical purposes we
can use \( Z=1 \) and \( M=0 \). 

Results are similar at \( \beta =5.7 \) though there one needs three
levels of HYP blocking to reduce M to 0 and Z to 1 as is illustrated
in Table \ref{quenched_results_b5.7}. Here the topological susceptibility
was measured on 350 \( 8^{3}24 \) lattices and the renormalization
constants are obtained from 500 heated \( 8^{4} \) lattices. The
predictions for the continuum topological susceptibility are consistent
from all three operators, \( \chi ^{1/4}=188(9) \)MeV. Within statistical
errors it is also consistent, though about 15\% lower, then the value
obtained on the \( \beta =6.0 \) data set. Whether the difference
is only statistical fluctuation or the \( a\approx 0.17 \)fm lattice
spacing at \( \beta =5.7 \) is too coarse to support all the topological
objects on the lattice cannot be decided with the present statistics.

\section{The Topological Susceptibility on dynamical configurations}

\subsection{Two flavors of staggered fermions }

We have analyzed several two-flavor configuration sets using HYP1-HYP3
operators. The renormalization factors have to be calculated independently
at every parameter value. The results for both Z and M were very similar
to the quenched values at similar lattice spacing. We found that on
configurations with lattice spacing \( a\approx 0.1 \)fm the renormalization
constants of the HYP2 operator could be neglected, while configurations
with \( a\approx 0.17 \)fm the HYP2 operator had a few percent correction
from the renormalization constants, the HYP3 operator had none.
\begin{table}
{\centering \begin{tabular}{|c|c|c|c|c|c|c|c|}
\hline 
&
\( n_{conf} \)&
\( r_{0}/a \)&
\( a \){[}fm{]}&
\( am_{\pi } \)&
\( a^{4}\chi \times 10^{5} \)&
\( (r_{0}m_{\pi })^{2} \)&
\( \chi r_{0}^{4} \)\\
\hline
\hline 
\( \beta =5.7,am_{q}=0.01 \)&
83&
6.29(8)&
0.08&
0.252(2)&
2.6(3)&
2.5(1)&
0.041(6)\\
\hline 
\( \beta =5.7,am_{q}=0.015 \)&
46&
6.02(8)&
0.08&
0.293(2)&
3.5(6)&
3.1(1)&
0.046(10)\\
\hline 
\( \beta =5.7,am_{q}=0.025 \)&
33&
5.8(1)&
0.09&
0.388(1)&
5.8(9)&
5.1(2)&
0.065(15)\\
\hline 
\( \beta =5.415,am_{q}=0.025 \)&
201&
2.96(3)&
0.17&
0.4454(2)&
84(8)&
1.74(1)&
0.064(9)\\
\hline
\end{tabular}\par}

\caption{\label{dynamical_results_nf2}Results for the topological susceptibility
on \protect\( n_{f}=2\protect \) configurations. All lattices are
\protect\( 16^{3}32\protect \) standard thin link staggered fermion
configurations.}
\end{table}
Table \ref{dynamical_results_nf2} collects our results. All four
data sets are from the NERSC archive, the first three were generated
by the Columbia group, the last one by MILC\cite{Brown:1991qw-NERSC_paper, Bernard:1998ni-MILC_staggered_nf2_spectrum}.
All lattices are \( 16^{3}32 \) and use standard thin link staggered
fermions with plaquette Wilson gauge action. \( am_{\pi } \) was
measured in the original studies, \( r_{0}/a \) for the MILC lattice
is from Ref. \cite{Tamhankar:1999ce-MILC_nf=2_scale}. For the Columbia
lattices we have measured \( r_{0}/a \) using HYP blocked Wilson
loops. The HYP potential has greatly reduced statistical errors making
it possible to obtain reliable values even from 33-83 configurations\cite{Hasenfratz:2001hp-HYP-blocking}.
The lattice spacing in the fourth column was obtained using \( r_{0}=0.5 \)fm
and is listed for future reference. Since the different HYP topological
charge operators give consistent results, only the continuum value
\( a^{4}\chi  \) is listed. The topological susceptibility for the
first data set has been measured in Ref. \cite{Hasenfratz:1999ng-spatial-correlation}
using 10-40 APE smeared operators. The values for \( a^{4}\chi  \)
with APE smeared operators are consistent with the present result.
\begin{figure}
{\centering \resizebox*{14cm}{!}{\includegraphics{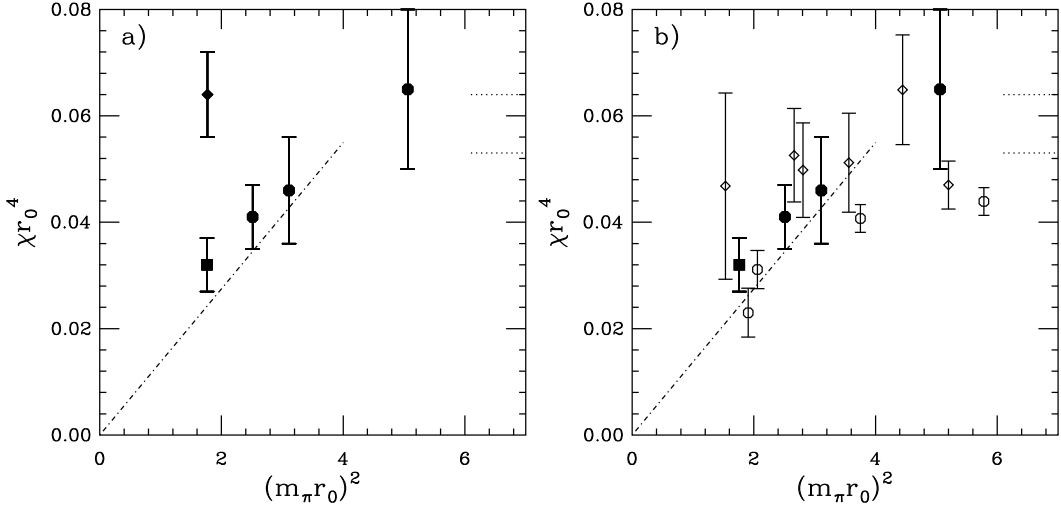}} \par}

\caption{\label{chi_vs_mpi_nf2}\protect\( \chi r_{0}^{4}\protect \) as the
function of \protect\( (m_{\pi }r_{0})^{2}\protect \) with \protect\( n_{f}=2\protect \)
staggered fermions. The dashed line is the leading order theoretical
prediction from eqn.(\ref{chi dynamical}) using \protect\( f_{\pi }=92\protect \)MeV.
The dotted lines on the right indicate the quenched value of \protect\( \chi r_{0}^{4}\protect \).
a) Staggered fermions. Filled octagons: Columbia data sets; filled
diamond: MILC data set; filled square: MILC\cite{DeTar}. b) Same
as a) with Wilson/clover data added. Octagons: UKQCD; diamonds: SESAM/T\protect\( \chi \protect \)L. }
\end{figure}
In Figure \ref{chi_vs_mpi_nf2}/a \( \chi r_{0}^{4} \) is plotted
as the function of \( (m_{\pi }r_{0})^{2} \). The filled octagons
correspond to the Columbia data sets with \( a\leq 0.1 \)fm. The
filled diamond is the MILC lattice data with \( a\approx 0.17 \)fm.
The filled square at the lowest \( m_{\pi }r_{0} \) value is also
from MILC. It is on a \( 24^{3}64 \), \( \beta =5.6 \), \( am_{q}=0.01 \)
configuration set and was measured by C. DeTar using 10-20 level APE
blocked topological operators\cite{DeTar}. The lattice spacing on
these configurations is \( a\approx 0.1 \)fm, \( r_{0}/a=4.99 \).
The dashed line in Figure \ref{chi_vs_mpi_nf2} is the leading order
theoretical prediction from eqn.(\ref{chi dynamical}) using \( f_{\pi }=92 \)MeV.
The three lowest \( m_{\pi }r_{0} \) data points on the finer lattices
with \( a\leq 0.1 \)fm are consistent, though a bit higher then the
leading order theoretical curve. However, the data point with \( a\approx 0.17 \)fm
is very different. The topological susceptibility on those lattices
is consistent with the quenched result even though the pion mass is
fairly small. Since on quenched configurations the topological susceptibility
can be measured successfully even on coarse lattices, this discrepancy
is not likely to be the consequence of the gauge system. Rather, it
appears that the instantons do not feel the effect of light staggered
fermions on coarse lattices. Flavor symmetry violation of staggered
fermions can explain these findings. In Figure \ref{chi_vs_mpi_nf2}/b
published Wilson/clover data is added to the \( a\leq 0.1 \)fm staggered
data. Octagons correspond to the clover fermion simulations of UKQCD,
diamonds to the Wilson fermion simulations of SESAM/T\( \chi  \)L.
All Wilson/clover data has lattice spacing \( a\leq 0.1 \)fm. The
data from UKQCD is consistent with the staggered results and with
the leading order theoretical curve as well. However the topological
susceptibility with unimproved Wilson fermions does not follow the
expected behavior at small pion mass. It remains large, consistent
with the quenched result, the same behavior we saw with staggered
fermions at lattice spacing \( a\approx 0.17 \)fm. It appears that
Wilson fermions have a very different effect on instantons even on
finer, \( a\approx 0.1 \)fm lattices than clover or even unimproved
staggered fermions. This is likely the consequence of chiral symmetry
breaking of the Wilson action.

\subsection{Four flavors of staggered fermions }

\begin{table}
{\centering \begin{tabular}{|c|c|c|c|c|c|c|c|}
\hline 
&
\( n_{conf} \)&
\( r_{0}/a \)&
\( a \){[}fm{]}&
\( am_{\pi } \)&
\( a^{4}\chi \times 10^{4} \)&
\( (r_{0}m_{\pi })^{2} \)&
\( \chi r_{0}^{4} \)\\
\hline
\hline 
\( \beta =5.2,am_{q}=0.06 \), thin&
188&
2.85(2)&
0.18&
0.661(1)&
8.5(9)&
3.5(1)&
0.056(8)\\
\hline 
\( \beta =5.25,am_{q}=0.06 \), thin&
189&
3.48(3)&
0.14&
0.664(1)&
4.8(5)&
5.3(1)&
0.070(10)\\
\hline 
\( \beta =5.2,am_{q}=0.1 \),fat&
140&
2.97(3)&
0.17&
0.695(4)&
3.5(4)&
4.2(2)&
0.026(4)\\
\hline
\end{tabular}\par}

\caption{\label{dynamical_results_nf4}Results for the topological susceptibility
on \protect\( n_{f}=4\protect \) configurations. All lattices are
\protect\( 8^{3}24\protect \). The first two sets were generated
with standard thin link staggered fermions, the third one with N=3
APE blocked fat link fermions.}
\end{table}
Unfortunately we could not find any large \( n_{f}=4 \) staggered
data sets to use. In Ref. \cite{Knechtli:2000ku-dynamical_fatlinks},
where we proposed a dynamical fat link fermion update, we generated
three \( 8^{3}24 \) \( n_{f}=4 \) data sets to study flavor symmetry
breaking of thin and fat link actions. The first two sets were generated
with standard thin link staggered fermions, the third one with N=3
APE blocked fat link fermions. All three sets have lattice spacing
\( a\approx 0.17 \)fm as shown in Table \ref{dynamical_results_nf4}.
The pion mass values in the table are also from Ref.\cite{Knechtli:2000ku-dynamical_fatlinks}
but we have re-analyzed the potential data using HYP Wilson loops
and the \( r_{0}/a \) values listed in Table \ref{dynamical_results_nf4}
are more reliable and slightly different from the ones used in Ref.
\cite{Knechtli:2000ku-dynamical_fatlinks} 
\begin{figure}
{\centering \resizebox*{7cm}{!}{\includegraphics{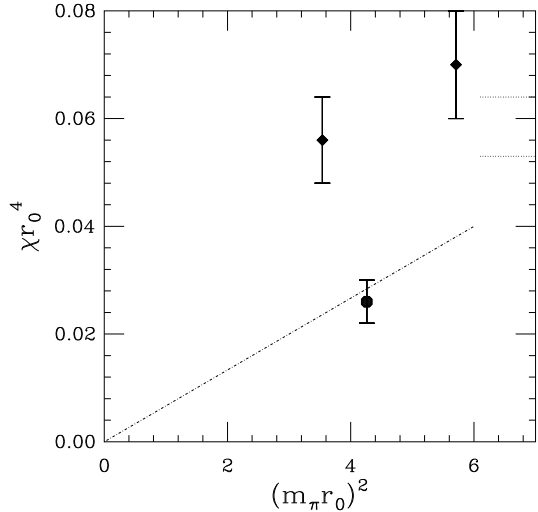}} \par}

\caption{\label{chi_vs_mpi_nf4}\protect\( \chi r_{0}^{4}\protect \) as the
function of \protect\( (m_{\pi }r_{0})^{2}\protect \) with \protect\( n_{f}=4\protect \)
staggered fermions. The filled diamonds are from thin link staggered
configurations, the filled octagon is from fat link dynamical configurations.
The dashed line is the leading order theoretical prediction from eqn.(\ref{chi dynamical})
using the experimental value \protect\( f_{\pi }=92\protect \)MeV.
The dotted lines on the right indicate the quenched value of \protect\( \chi r_{0}^{4}\protect \).}
\end{figure}
The difference in the topological susceptibilities between the thin
and fat link actions is striking. There is no Wilson/clover fermion
data available at \( n_{f}=4 \) to compare the staggered result.
In Figure \ref{chi_vs_mpi_nf4} we plot \( \chi r_{0}^{4} \) as the
function of \( (m_{\pi }r_{0})^{2} \) for the three data sets from
Table \ref{dynamical_results_nf4}. The dashed line is the leading
order theoretical prediction from eqn.(\ref{chi dynamical}) using
the experimental value \( f_{\pi }=92 \)MeV. The fat link action
data shows the expected behavior while the thin link action topological
susceptibility values are consistent with the quenched result, independent
of the dynamical quark mass. For the thin link action this is the
same behavior we saw with the \( n_{f}=2 \) data at similar lattice
spacing.

\section{Discussion and Summary}

The interaction between light quarks and instantons changes the QCD
vacuum substantially. One of the easiest way to get information about
this effect is through the topological susceptibility that is expected
to scale with the square of the pion mass in the small quark mass
limit.  

Our results indicate that thin link staggered fermions at lattice
spacing \( a\geq 0.17 \)fm do not have the expected continuum effect
on instantons, a lattice spacing \( a\approx 0.1 \)fm is needed to
recover the proper chiral behavior. In contrast to that, fat link
staggered fermions show the expected behavior even on coarser \( a\approx 0.17 \)fm
lattices. Can this be understood as the effect of flavor symmetry
violation? In Refs. \cite{Knechtli:2000ku-dynamical_fatlinks, Hasenfratz:2001hp-HYP-blocking}
we studied flavor symmetry violation on quenched lattices both with
the thin link, N=3 APE blocked and hypercubic blocked fat link fermions.
The two fat link formulations have about the same level of flavor
symmetry violations as measured by the relative mass splitting \( \Delta _{\pi }=(m_{\pi }-m_{G})/m_{G} \)
between the Goldstone pion \( m_{G} \) and the other pion like objects,
\( m_{\pi } \). For thin link fermions at \( (m_{\pi }r_{0})^{2}\approx 2.0 \)
we found \( \Delta _{\pi }\approx 0.7 \) at \( a\approx 0.17 \)fm,
\( \Delta _{\pi }\approx 0.2 \) at \( a\approx 0.1 \)fm for the
lightest non-Goldstone pions. The corresponding values for fat link
fermions are \( \Delta _{\pi }\approx 0.09 \) and \( \Delta _{\pi }\approx 0.01 \),
respectively. QCD with two(four) light flavors should have 3(15) light
pions. Apparently when there is only one light pion and the other
pion-like objects are 70\% or more heavier than the Goldstone pion,
the vacuum does not look like a two(four)-flavor QCD vacuum. The flavor
symmetry breaking has to be reduced below 20\% to get acceptable results.
Fat link fermions can do that easily even on coarse lattices. 

Available lattice data for Wilson/cover fermions can be understood
similarly. The topological susceptibility with clover fermions are
consistent with the theoretical predictions at lattice spacing \( a\approx 0.1 \)fm
but the Wilson fermion data at the same lattice spacing indicates
that unimproved Wilson fermions, that have much larger chiral symmetry
violations, do not have the expected continuum like interaction even
at \( a\approx 0.1 \)fm. Results from CP-PACS at lattice spacing
\( a\approx 0.17 \)fm indicate that at that lattice spacing not even
clover fermions can reproduce the continuum topological behavior.
Fat link clover fermions improve chiral symmetry and could provide
a better alternative to thin link clover fermions. It would be interesting
to find a parameter similar to \( \Delta _{\pi } \) that correlates
with the level of chiral symmetry breaking for the Wilson like fermions
and compare it with the topological susceptibility. 

One should be concerned about dynamical simulations where the topological
susceptibility is not reproduced correctly, since that indicates that
the vacuum at those simulations is more like the quenched vacuum rather
than the expected dynamical one. Improving chiral symmetry can have
a profound effect. This point is further underscored in a forthcoming
publication about the finite temperature phase diagram obtained with
fat link fermions\cite{Hyp_thermo}.

\begin{acknowledgments}
This work was largely inspired by the collaboration with F. Knechtli
on fat link fermions and hypercubic blocking. G. Bali convinced me
that in order to compare the topological susceptibilities from different
simulations one needs to use a reliable scale, like \( r_{0} \).
That prompted the potential measurements with hypercubic blocked links.
S. Gottlieb helped me out with some of the MILC staggered fermion
data. The numerical calculations of this work were performed on the
Colorado-HEP  \( \alpha  \) cluster. 
\end{acknowledgments}
\bibliographystyle{apsrev.bst}
\bibliography{hypt}

\end{document}